\documentstyle[a4,11pt,psfig]{article}
\def\etal{{et~al.}\ }
\def\be{\begin{equation}}
\def\ee{\end{equation}}
\def\ie{{{\frenchspacing i.{\thinspace}e. }}}

\def\la{\mathrel{\hbox{\rlap{\hbox{\hbox{$_\sim$}}}\raise2pt\hbox{$<$}}}}
\def\ga{\mathrel{\hbox{\rlap{\hbox{\hbox{$_\sim$}}}\raise2pt\hbox{$>$}}}}
\def\G{\Gamma}
\def\a{\alpha}
\def\normP{{\cal{P}}}

\begin{document}

\title{Circular scans for CMB anisotropy observation and analysis}
\author{J.~Delabrouille,$^{1,2}$ K.M. ~G\'orski$^{3,\,4}$ and E.~Hivon$^3$}

\maketitle

\begin{center}
{\small
\noindent
$^1$ Institut d'Astrophysique Spatiale, CNRS \& Universit\'e Paris XI, 
b\^at 121, 91405 Orsay Cedex, France

\noindent
$^2$ Enrico Fermi Institute, University of Chicago, 5460 South Ellis Avenue, 
Chicago, IL 60510, USA

\noindent
$^3$ Theoretical Astrophysics Center, Juliane Maries Vej 30, DK-2100, 
Copenhagen, Denmark

\noindent 
$^4$ On leave from Warsaw University Observatory, Warsaw, Poland} 
\end{center}

\begin{abstract}
A number of experiments for measuring anisotropies 
of the Cosmic Microwave Background use
scanning strategies in which temperature
fluctuations are measured along circular scans on the sky. 
It is possible, from a large number of such intersecting circular scans, 
to build two-dimensional sky
maps for subsequent analysis. However, since instrumental
effects --- especially the excess low-frequency $1/f$ noise --- 
project onto such two-dimensional  maps in a non-trivial way,
we discuss the analysis approach
which focuses on information contained in 
the individual circular scans. 
This natural way of looking at CMB data from experiments scanning on the
circles combines the advantages of
elegant simplicity of Fourier series for the computation of 
statistics useful for constraining cosmological scenarios,
and
superior efficiency in analysing and quantifying
most of the crucial  instrumental effects.

\end{abstract}


\section{Introduction}

The exploration of Cosmic Microwave Background (CMB)  is undergoing 
a boom in  both theoretical and experimental directions. The
planning  of two ambitious space missions, MAP and PLANCK, 
to be launched in
the beginning of the XXI-st century, is one of the sources of 
stimulus for research in this field. In addition, several long duration 
balloon-borne experiments are scheduled for operation within the next few years
time. 

Many of the future CMB
anisotropy experiments will perform circular scans on the sky.
MAP\footnote{MAP home page: http://map.gsfc.nasa.gov/}, 
PLANCK\footnote{Planck home page:
http://astro.estec.esa.nl/SA-general/Projects/Cobras/cobras.html}
(Bersanelli et al., 1996), and  the balloon-borne experiments 
TopHat\footnote{MSAM/TopHat home page:
http://cobi.gsfc.nasa.gov/msam-tophat.html} and 
BEAST (Lubin, private communication), 
will collect data from a large 
number of intersecting circles, which will then be merged into
two-dimensional sky maps. 
Smaller ground-based experiments as DIABOLO 
(Beno\^{\i}t et al., in preparation) 
may scan on only a few circles. 
In winter 1997 for instance, a one-ring observation of anisotropies was 
done at the POM2 telescope in the french alps with the DIABOLO instrument. 

A common problem in the analysis of the data obtained by performing
one dimensional
sky scans 
is that all instrumental effects (other than 
smoothing by the beam) occur in the monotonic time domain of the 
data stream, but need to be projected onto the often complicated 
geometry of directions of observation, which is essential for 
the interpretation of the results.
Excess low-frequency (typically $\sim 1/f$) noise, for instance, 
is an instrumental 
effect that is not easily analysed when projected onto two-dimensional maps
(Delabrouille, 1997). 

In the following,
we explore  the relation between the Fourier 
spectrum of temperature fluctuations measured on a circle
and spherical harmonic expansion 
coefficients on the sphere, with associated uncertainties in the framework
of gaussian statistics, and investigate how circular scans 
permit to analyse naturally both the signal of cosmological origin and 
some of the 
effects coming from the instrument.

\section{CMB anisotropies on circular scans}

\subsection{Fourier spectra}

The usual expansion in spherical harmonics of the statistically
isotropic CMB 
temperature fluctuations on the sky observed with a symmetric 
beam reads:

\begin{equation}
	T(\vec{n}) = \sum_{\ell=1}^{\infty} \sum_{m=-\ell}^{\ell} a_{\ell m}\, B_\ell\, 
	Y_{\ell m}(\vec{n}).
\label{eq:Ylm-sky}
\end{equation}
Here, the coefficients $a_{\ell m}$ are assumed to be zero-mean 
Gaussian deviates with variances  given by 
$\langle \vert a_{\ell m}\vert^2 \rangle = C_\ell$
(on assumption of statistical isotropy of the CMB temperature perturbations,
here expressed via the ensemble averaging),
$B_\ell$ is the  beam response function 
(we have assumed for simplicity that the beam is symmetric),
and 
a unit vector $\vec{n}$ defines the position on the sphere. 
The functions $Y_{\ell m}(\vec{n}) = Y_{\ell m}(\theta, \phi)$
are the orthonormal spherical harmonics, defined, 
for $-\ell \le m \le\ell$, as 
\begin{eqnarray}
	Y_{\ell m}(\theta, \phi) & = & \normP_\ell^m(\theta)\ e^{im\phi},
\end{eqnarray}
with
\begin{eqnarray}
	\normP_{\ell m}(\theta) & = &
	\sqrt{\frac{2\ell+1}{4\pi}\frac{(\ell-m)!}{(\ell+m)!}}
	\ P_{\ell m}(\cos\theta), \quad{\rm for\ } m\ge 0, \nonumber\\
	& = & (-1)^{|m|} \normP_{\ell|m|}(\theta),\quad{\rm for\ } m< 0, \nonumber
\end{eqnarray}
where $P_\ell^m$ are the associated Legendre polynomials. This definition
of the $Y_{\ell m}$, together with the fact that the temperature is real, 
implies $a_{\ell m} = (-1)^m a_{\ell -m}^*$.

If only one ring of angular radius ${{\Theta}}$ on the sky is being scanned, 
it is convenient to use coordinates such that the observed
circle is the set of points of the sphere at constant colatitude
${ {\Theta}}$. 
The CMB temperature on this circle, $T({ {\Theta}},\phi)$, 
can be decomposed uniquely
in the form of a Fourier series:
\begin{equation}
	\alpha_m = \frac{1}{2\pi}\int_0^{2\pi}\; d\phi\; T({{\Theta}},\phi)\;
	e^{-i m\phi} \; = 
	\sum_{\ell=|m|}^{\infty} a_{\ell m}\, B_\ell\, 
 	\normP_{\ell m}({ {\Theta}}).
\label{eq:Fourier-components}
\end{equation}

If  $a_{\ell m}$ are uncorrelated  gaussian random variables, as expected 
in an inflationary scenario (Bardeen et al., 1986), the same is the case also 
for the ring mode amplitudes $\alpha_m$.
The corresponding Fourier spectrum, whose components we shall denote
as $\Gamma_m$, is the one dimensional analog of the $C_\ell$ spectrum on the
sky. The anisotropy Fourier spectrum $\Gamma_m$ is obtained from the $C_\ell$
spectrum by:
\begin{equation}
	\langle \alpha_m \alpha_{m'}^* \rangle = \Gamma_m \delta_{mm'},
\end{equation}
and
\begin{equation}
	\Gamma_m = \sum_{\ell=|m|}^{\infty} C_\ell\, B_\ell^2
 	\normP_{\ell m}^2({{\Theta}}).
\label{eq:lien_spectres}
\end{equation}
This last equation allows a straightforward computation of the
ring anisotropy power spectrum  $\Gamma_m$  given the instrumental 
specifications 
($\Theta$ and $B_\ell$)
and the cosmological model  
($C_\ell$ spectrum as a function of the relevant parameters).

\begin{figure}[ht]
\hbox{
 \psfig{figure=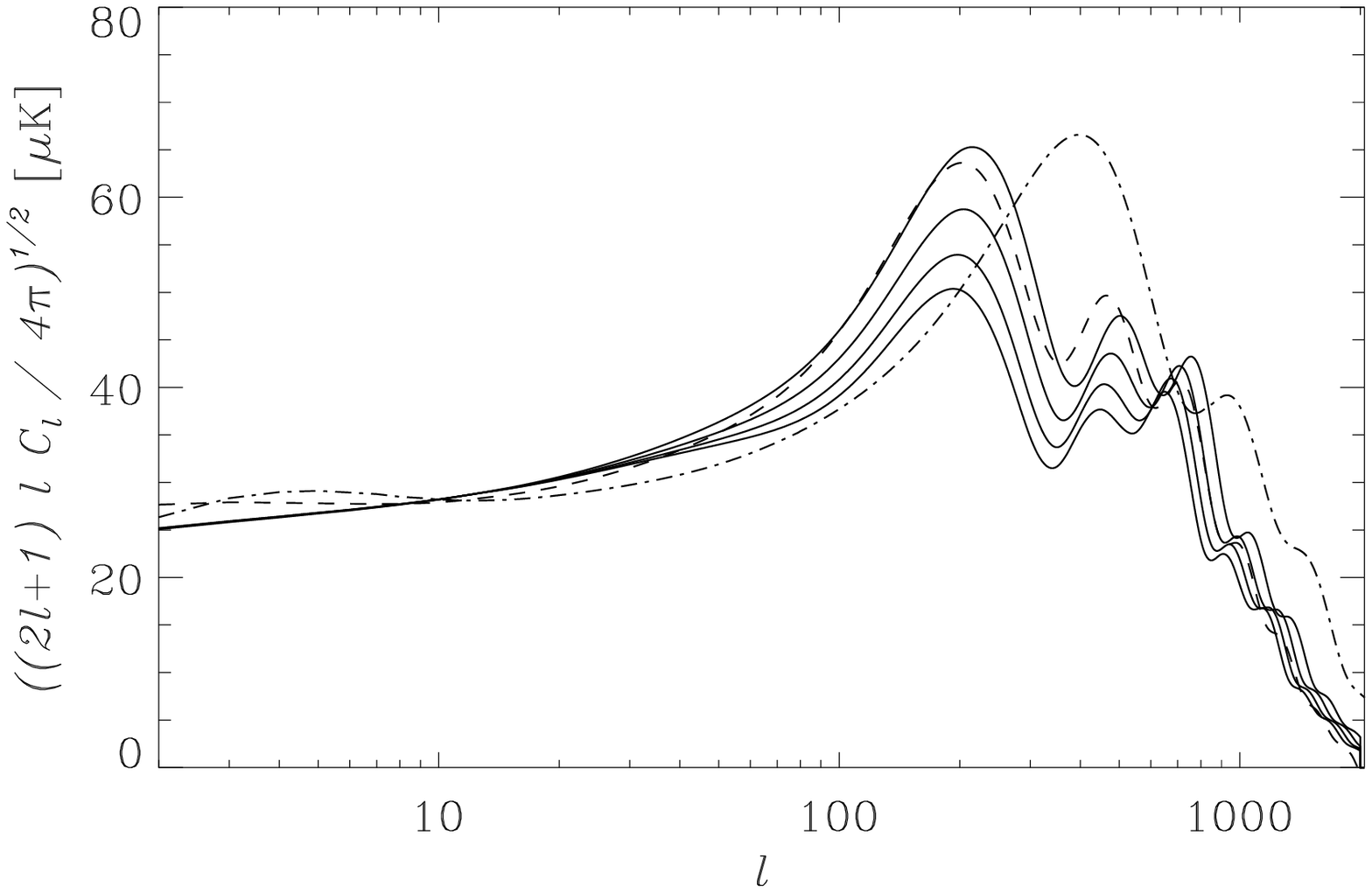,width=.49\textwidth}
 \psfig{figure=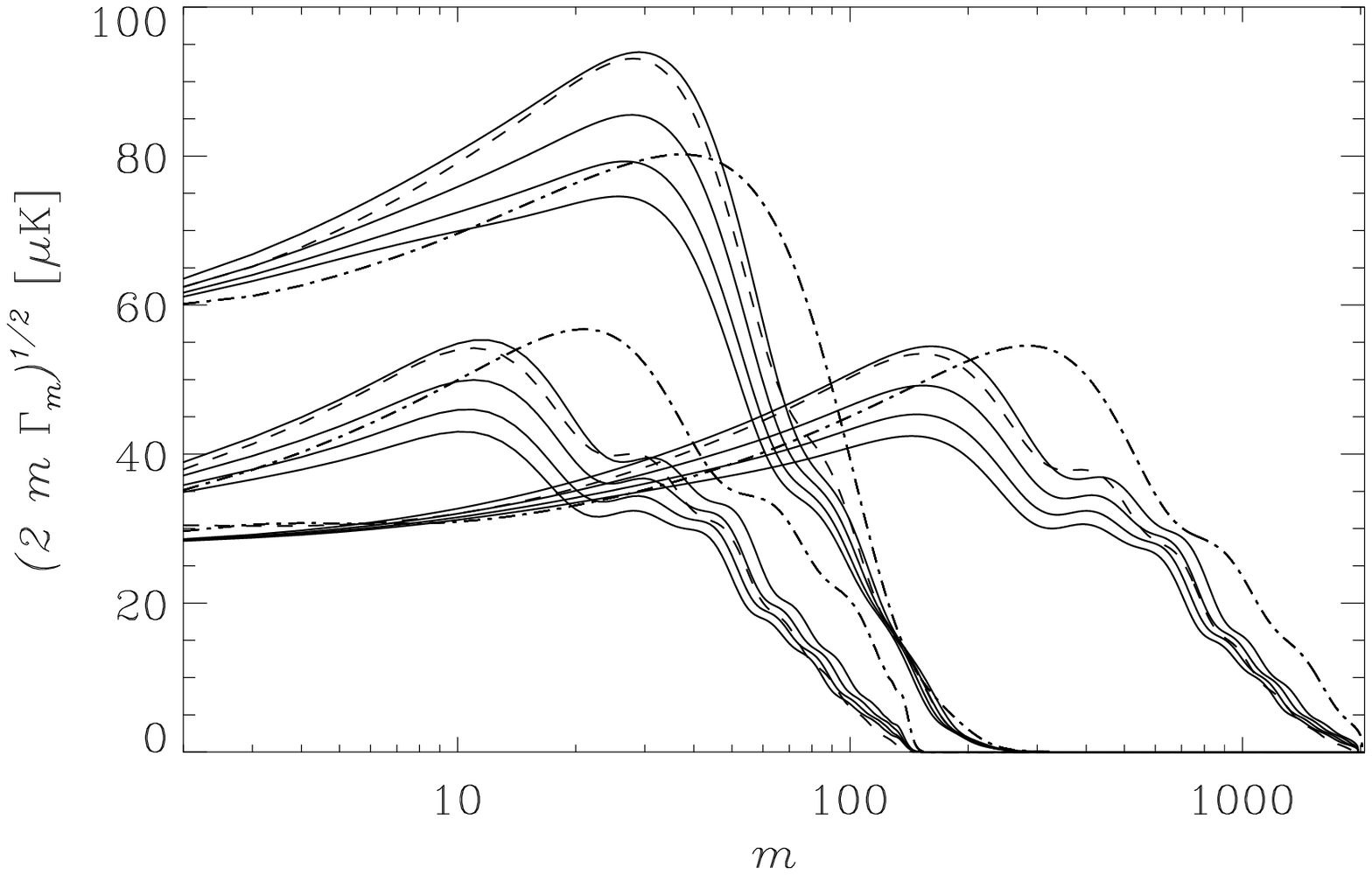,width=.49\textwidth}
}
\caption[]{Left panel: Full sky power spectra $C_\ell$ of CMB anisotropy,
smoothed with a gaussian beam of FWHM = 7 arcmin,
in various CDM cosmological models normalized roughly to match the
CMB anisotropy detected by {\it COBE}-DMR.
All models obey the primordial nucleosynthesis constraint  $\Omega_b h^2
= 0.015$. 
Solid lines (top to bottom) correspond to flat, matter dominated models
with Hubble constant $h=0.5,\, 0.6,\, 0.7,\, 0.8$.
The dashed line  shows a cosmological constant dominated model 
with $\lambda=0.7$ and $h=0.8$, and the dot-dashed line shows
an open model with $\Omega_0=0.3$ and $h=0.65$.
Right panel: One-dimensional  power spectra of CMB anisotropy
on circular scans computed for the same models as shown in the left panel. 
Three groups of curves as viewed from left to right correspond to different 
specifications (by the angular radius of the circle, $\Theta$,
and the FWHM of the beam)
of the geometry of observations:
the left group corresponds to
a small ring of $\Theta=4^\circ,\, {\rm FWHM} =5'$;
the middle group (amplitudes multiplied by 2 to avoid overlap with the
other curves) corresponds to an intermediate size ring of 
$\Theta = 12^\circ,\, {\rm FWHM} = 20'$; finally, 
the right group corresponds to a large ring of 
$\Theta = 80^\circ,\, {\rm FWHM} = 7'$.
Both the left and right panels show the CMB anisotropy rms amplitude per
logarithmic increment of the relevant index $\ell$ or $m$, 
taking into account the number of
degrees of freedom per mode (factors $(2\ell +1)\ell$, and $2m$, respectively).}
\label{fig:Cl+Gamma}
\end{figure}

Figure \ref{fig:Cl+Gamma} illustrates some properties of the ring spectrum $\Gamma_m$ 
as compared to the usual full sky spectrum $C_\ell$.
Observational configurations chosen for the plot correspond to  1)
a single ring scan
of the PLANCK satellite observing with the High Frequency Instrument 
(the large ring), 
2) a single ring of the TopHat balloon experiment (the intermediate ring), and
3) a hypothetical small ring and a  narrow beam configuration. 
One should note that the ring anisotropy power spectra preserve the dependence
on cosmological parameters observed by the $C_\ell$ curves, and, 
if the resolution of the instrument is good enough and the ring size 
sufficiently large,
display clearly the array of adiabatic peaks. This promotes the strategy
of observing the CMB anisotropy on the rings to a very interesting status.  
Clearly one should be able to conduct a well guided analysis of such a data with
the aid
of a simple statistic, $\Gamma_m$, which, due to the natural geometric
reduction from the whole sky to the circle, captures all those features of 
theoretical
CMB anisotropy predictions --- direct reflection of physical processes
that perturb CMB temperature in the shape of power spectrum, and the
traceable dependence on cosmological parameters --- which created high level of
expectations for  the
full sky CMB mapping missions.  
 
\begin{figure}[ht]
\begin{center}
\hbox{
\hskip 1.5cm
\psfig{figure=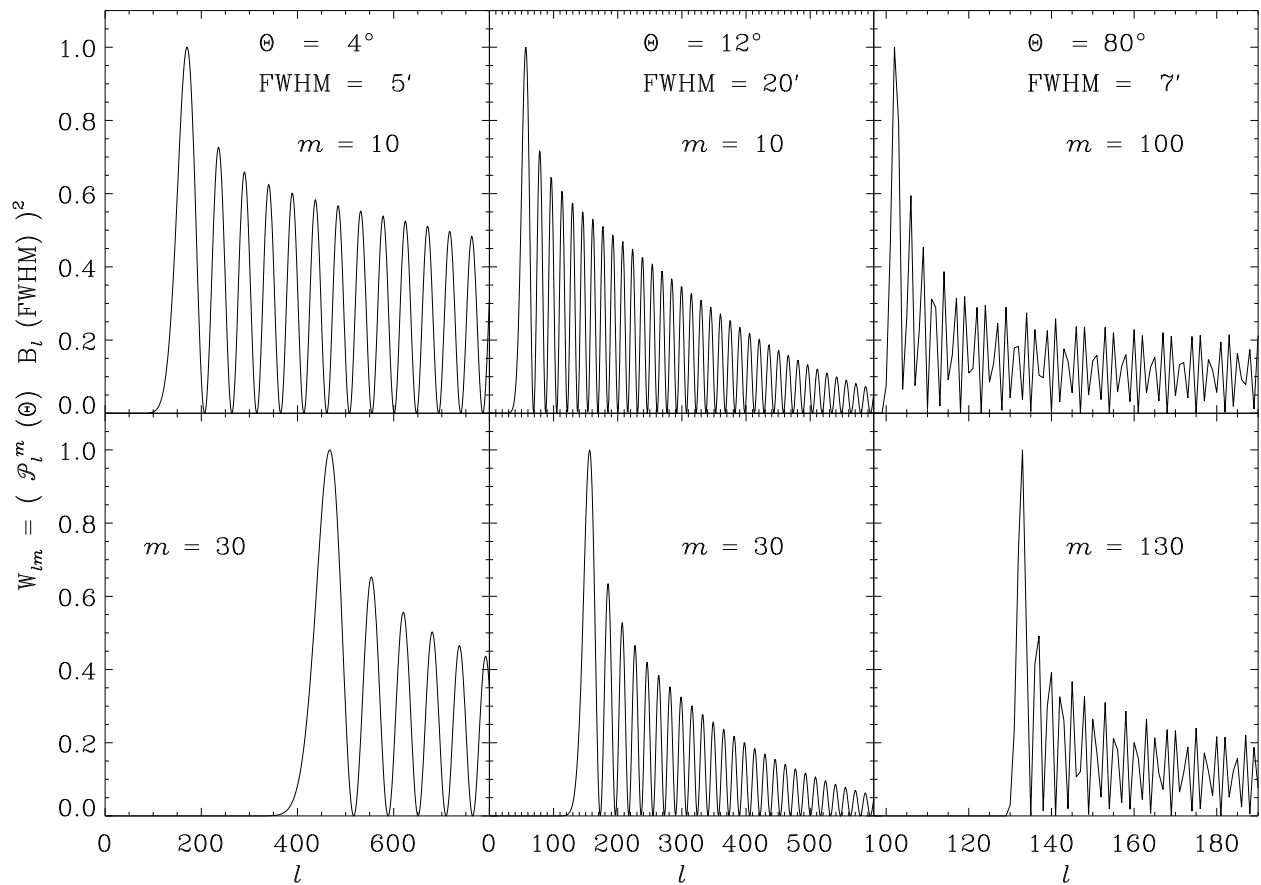,width=0.9\textwidth}
}
\end{center}
\caption[]{Window functions for computation of the CMB anisotropy
ring power spectrum
$\Gamma_m $, see Eqs. (\ref{eq:lien_spectres}), 
and (\ref{eq:window_fct}), normalised to
unity at  their maxima.}
\label{fig:window}
\end{figure}

Ring power spectrum  coefficients $\G_m$ can be viewed  as the estimators of the
spectrum $C_\ell$ integrated over $\ell$ with a certain $m$-dependent 
window function whose coefficients are simply:\\
\begin{equation}
	W_\ell^{(m)} =  {\normP_{\ell m}}^2( {\Theta})
	{B_\ell}^2.
	\label{eq:window_fct}
\end{equation}
A number of such window functions $W_\ell^{(m)}$, corresponding to
the ring configurations used in Fig. \ref{fig:Cl+Gamma}, are displayed in 
Fig. \ref{fig:window}. 
An important, if simple, property of these window functions, and, hence,
the  $\G_m$ coefficients, is that the power at mode $m$ is generated 
by only those components of sky anisotropy for which $\ell \ge m$.
This is rather important for the suborbital attempts to measure the CMB
anisotropy, which have to cope with the atmospheric effects.
It has been the case for a long time that due to the variation
from zenith to horizon of the length of path through the Earth's atmosphere,
and the associated  systematic effects in the measurements of CMB temperature,
the preferred scanning strategy in CMB experiments
was usually along the lines of constant elevation. Can one, in a suborbital
experiment, afford to
scan on any other circle than that at constant elevation? If the atmospheric 
temperature
gradient is predominantly large-scale compared to the size of the observing ring,
its contaminating effects will be confined to low-$m$ modes of the
ring anisotropy spectrum, and high-$m$ modes will be algebraically
decoupled from such a source of spurious anisotropy.  

To what extent would a measurement of CMB anisotropy on the ring allow us to
estimate the parameters of cosmological model used to describe the data?
To be able to answer this question we need first to clarify the statistical
properties of circular modes of CMB anisotropy, and the properties of 
instrumental noise as projected on the ring geometry of observations.
We shall see momentarily  that the estimation of uncertainties in 
measurements of $\alpha_m$
due to white noise, excess low-frequency noise, 
or instrumental systematic effects
is simpler and  more secure in the case of ring anisotropy 
than in the case of two dimensional maps used in  $C_\ell$ estimation.

\subsection{Statistics of Fourier transform of CMB anisotropy measurements on 
the rings}
Here we address the issue of cosmic variance of the ring modes of the CMB
anisotropy.
Fourier coefficients of temperature fluctuations measured on the circle,
$\alpha_m$-s, 
are linear
combinations of $a_{\ell m}$-s, whose statistical properties they hence
inherit. Specifically,  Gaussian distribution of $a_{\ell m}$-s renders
Gaussian distribution of $\a_m$-s. Statistical isotropy of the model CMB 
anisotropy field results in a very special property of the $m$-independence
of variance --- $\langle\vert a_{\ell m}\vert ^2\rangle = C_\ell$, which makes
the $C_\ell$ spectral coefficients $\chi_{(2\ell +1)}^2$ distributed.
In the case of Fourier coefficients of the ring data the number of degrees 
of freedom associated with a specific $m$ value is reduced to only two ---
$\pm m$ waves have identical variance. Hence, the cosmic variance on the
$\G_m$
coefficients is described by the $\chi_2^2$ probability
distribution, independent of the value of $m$.

\begin{figure}[ht]
\begin{center}
\vbox{
\hbox{\hskip 1.7cm
\psfig{figure=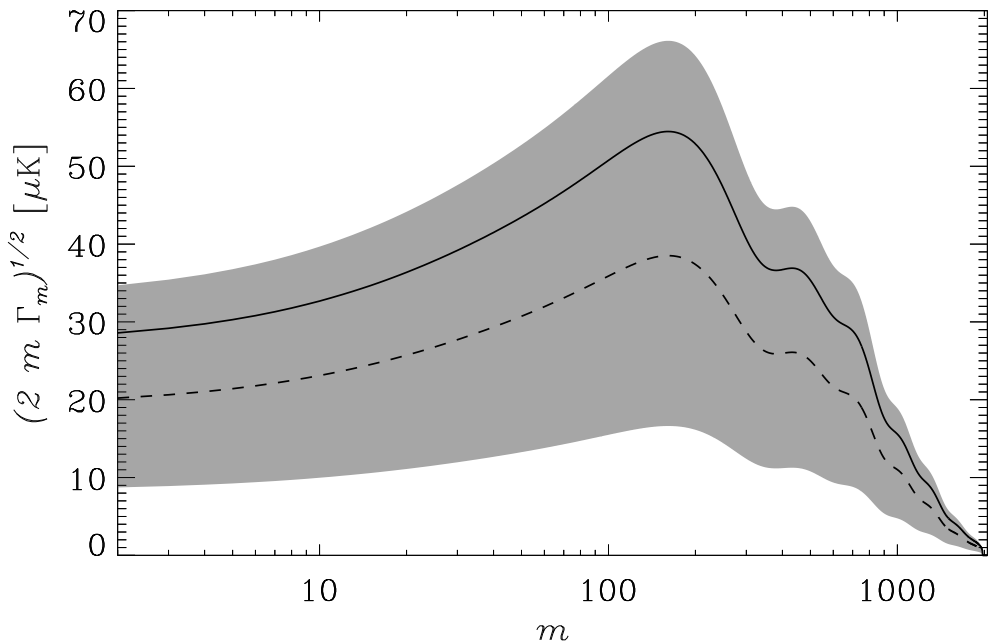,width=0.8\textwidth}
}
\vskip -1.5cm
}
\end{center}
\caption[]{Statistics of the rms power of the CMB anisotropy
measured on the ring. Gray band covers the 68\% confidence range for each
rms power coefficient and scales identically at each value of $m$ in terms
of the variance, which is shown by solid line. Dashed line shows the loci
of the most probable values of the rms power, illustrating the skewness of the 
distribution.}
\label{fig:error}
\end{figure}
 
In Figure  \ref{fig:error} we illustrate the statistical properties 
of ring Fourier coefficients of CMB anisotropy. The plot shows 
a few features of probability distribution of the rms power
in mode $m$. It is clear that the cosmic variance in the ring Fourier 
coefficients is large, and, unlike in the case of full sky measurements
of CMB anisotropy, when available at  small enough angular scales,
the statistics of power at high $m$ do not improve compared to those in low-$m$
modes, as the ring configuration allows just two degrees of freedom  per 
mode, independent of $m$. It is clear that the only opportunity for improvement
of statistics of the measurements of ring modes of CMB anisotropy
rests with repeating the observations on many rings.
The essential issues of assessing the ring to ring covariance of the
Fourier modes of CMB anisotropy will be addressed in detail
in a separate paper (G\'orski, Delabrouille, and Hivon, in preparation).
 
\section{Noise on circular scans}

Now we turn to the estimation of the projection of instrumental noise onto
a ring of data. Noise, unlike the CMB temperature anisotropies,
is a time-dependent process which is independent (to first order) of the 
position on the sky. The same is true of many instrumental effects in the
measured 
signal (system impulse response, sampling,...), but we first limit our 
discussion
to a simple Gaussian noise as an illustration of the connection between 
the measured data stream and the signal reprojected on the ring.

Let us assume that data is collected by a total-power experiment as
PLANCK, TopHat, BEAST or DIABOLO, and that a circle on the sky
is scanned continuously $N$ times, 
with a spinning frequency $f_{\rm spin}$ (spinning period $T_{\rm spin}$). 
The output of each detector is a signal $s(t)$, whose value as a function 
of time can be written as:\\
\begin{equation}
s(t) = u\left(\phi(t)\right) + n(t).
\label{eq:sn}
\end{equation}
In this equation, $\phi(t)=2 \pi f_{\rm spin} t$ is the longitude angle 
on the circle. The
way it depends upon time reflects the scanning strategy. 
$u(\phi) = T(\Theta,\phi)$ is the useful signal from the sky, 
\ie the temperature anisotropy smoothed with the beam.
It is a $2 \pi$-periodic function of $\phi$ for fixed $\Theta$.
If anisotropies are correctly represented by a gaussian random field, the 
Fourier components $u_m$ of the Fourier series decomposition of $u(\phi)$ 
are realisations of the random variables
$\a_m$ defined in the 
previous section. The noise $n(t)$ is assumed to be a realisation
of a Gaussian
stationary random process with zero mean and a bilateral 
spectrum ${\cal S}_n(f)$ which, over a large range of frequencies,
can be typically of the form:
\begin{equation}
{\cal S}_n(f) = a \left( 1+ \left( \frac{f_{\rm knee}}{|f|} \right)  \right)
\end{equation}
This is the simple case of the so-called $1/f$ noise.

Given the signal timeline as defined in Eq. (\ref{eq:sn}), 
the estimation of the Fourier components of anisotropy on the ring
can be done by computing the integral:
\begin{eqnarray}
s_m & = & \frac{1}{N T_{\rm spin}} \int_{0}^{N T_{\rm spin}} 
	s(t) \exp \left( -\frac{2i \pi m t}{T_{\rm spin}} \right) {\rm d}t \nonumber \\
    & = & u_m + \frac{1}{N T_{\rm spin}} \int_{0}^{N T_{\rm spin}} 
	n(t) \exp \left( -\frac{2i \pi m t}{T_{\rm spin}} \right) {\rm d}t.
\end{eqnarray}
Under the present hypotheses (no foregrounds, uncorrelated gaussian noise),
this quantity $s_m$ is an unbiased estimator of $\a_m$, which variance
originates in only two sources of uncertainty.
First, there is the theoretical ``cosmic" or ``sample" variance due to the fact
that $u_m$ is just one realisation of $\a_m$, as discussed in paragraph 2.2.
Second, there is the uncertainty due to the detection process (\ie due to
the noise $n(t)$).
Let us denote as $n_m$ the quantity
\begin{equation}
n_m = \frac{1}{N T_{\rm spin}} \int_{0}^{N T_{\rm spin}}
	n(t) \exp \left( -\frac{2i \pi m t}{T_{\rm spin}} \right) {\rm d}t. 
\end{equation}
Each of the $n_m$-s can be viewed as a realisation of a gaussian random
variable. They obey the correlation statistics:
\begin{equation}
\langle n_mn_{m'}^* \rangle = \int_{-\infty}^{\infty}
				{\cal S}_n(f) H_m(f) H_{m'}^*(f), \nonumber  {\rm d}f
\label{eq:noise-correlation}
\end{equation}
where $H_m(f)$ is given by:
\begin{equation}
H_m(f) = {\rm e}^{(i \pi N(f/f_{\rm spin}-m))} \times
\frac{\sin(\pi N(f/f_{\rm spin}-m))}{\pi N(f/f_{\rm spin}-m)}.
\end{equation}

Interestingly, but not surprisingly, there may be some correlations 
between Fourier amplitudes of
the noise reprojected on the ring, depending on the nature of the noise 
spectrum ${\cal S}_n(f)$. However, if
the number $N$ of scans on a circle is large and
the noise spectrum ${\cal S}_n$ is not too steep ($1/f$ is fine), the
expectation value of $n_mn_m^*$ becomes:
\begin{equation}
\langle n_{m}n_{m'}^* \rangle \simeq
\frac{{\cal S}_n(mf_{\rm spin})}{NT_{\rm spin}} \delta_{mm'}.
\label{eq:noise-level}
\end{equation}

Thus, for each individual
$m$ value, the uncertainty in $s_ms_{m}^*$ induced by noise depends on
$f_{\rm spin}$ through the noise spectrum value ${\cal S}_n(mf_{\rm spin})$,
and upon the total integration time $NT_{\rm spin}$. For a given experiment, 
the spinning frequency and total integration time should be optimised 
by comparing the level of the noise spectrum to the expected spectrum of the
signal of astrophysical origin. 
Usually, the noise spectrum decreases with increasing 
$f$, and thus the minimisation of $n_mn_m^*$ requires to spin fast, up to a 
compromise which depends on other experimental considerations.

If the number of scans on a circle, $N$, is large, and the noise spectrum 
${\cal S}_n$ is not too steep, the expectation value of $s_ms_m^*$ averaged 
over sky realisations of the CMB {\it and} noise realisations is:
\begin{equation}
\langle s_{m'}s_m^* \rangle \simeq \left( \G_m + 
\frac{{\cal S}_n(mf_{\rm spin})}{NT_{\rm spin}} \right) \delta_{mm'}.
\label{eq:signal-spectrum}
\end{equation}

\begin{figure}[htbp]
\psfig{figure=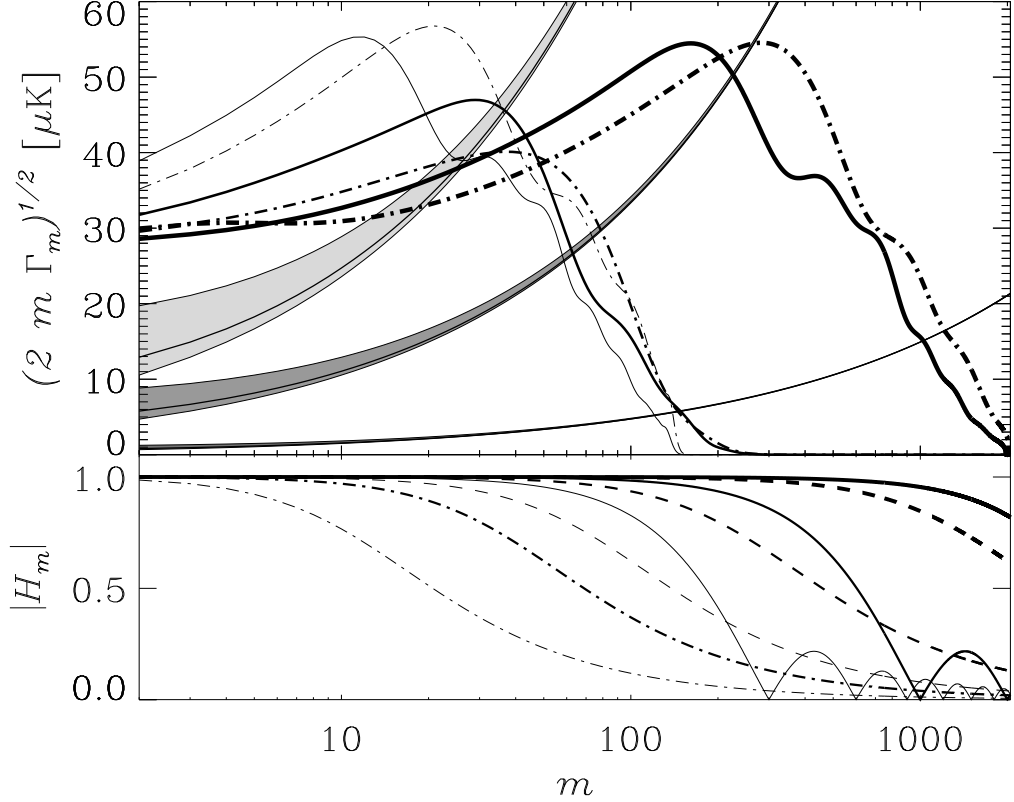,width=0.9\textwidth}
\caption[]{Top panel: Power spectra $\G_m$ of signal and noise 
for  the following set-ups of CMB anisotropy measurements:
1)  the ring radius of $\Theta = 4^\circ$ and FWHM = 5 arcmin --- 
light lines,
2) $\Theta = 12^\circ$ and FWHM = 20 arcmin --- medium-width lines, and
3) $\Theta = 80^\circ$ and FWHM = 7 arcmin --- heavy lines. 
Solid lines show the predictions of standard CDM model, and dot-dashed lines 
those of an $\Omega_0=0.3$ open model, both normalized to {\it COBE}-DMR. 
Three 
gray bands show the range of rms values of
instrumental noise contributions. Bottom edges of the gray bands
correspond to pure white noise behaviour of the detector. Solid lines
inside the 
bands illustrate the noise enhancement due to the $1/f$ component 
when the knee frequency 
is equal to the spin frequency on the ring scan.
The upper edges of the gray bands correspond to  $f_{\rm knee}/f_{\rm spin} =5$
(excess low frequency noise grows with $f_{\rm knee}/f_{\rm spin}$).
The assumed (white noise) detector  performance is characterized by 
sensitivity of 1 mK$\sqrt{s}$ and  total integration time of
20 and 100 hours for the 
upper two bands (light and medium gray respectively), while the
third, narrow, dark gray noise band corresponds to a 
sensitivity of 20$\mu$K$\sqrt{s}$ and integration time of 2 hrs.
Bottom panel: Illustration of the effects of sampling and
bolometer time constant (see text). 
Heavy lines correspond to spin period of 1 min, and illustrate the effect of 
sampling at 10 ms --- solid line, and bolometer time constant of 6 ms --- dashed 
line.
Medium-width  and light curves correspond to spin period of
10 s, and 3 s, respectively. 
For these two spin rates the solid lines illustrate the effect 
of sampling at 10 ms, and the  dashed and dot-dashed lines 
show the effect of bolometer time constant of 6 ms, and 40 ms, respectively.}
\label{fig:fig_G_m_noise}
\end{figure}

Thus a direct comparison of the signal and noise spectra for typical 
next-generation experiments is  quite straightforward in the ring formalism
and is presented here in  Fig. \ref{fig:fig_G_m_noise}.

In the top panel of Figure \ref{fig:fig_G_m_noise}, we have plotted together
the rms values of  $\G_m$ for three different
experiments and two different cosmological models, and the noise-induced 
sensitivity limits to individual $\G_m$ components. Experimental 
parameters chosen for the plot are representative  of the next
generation balloon-borne or satellite experiments. 
The pure  white noise contribution, proportional to $m^{1/2}$, has
the shape of an exponential because of the logarithmic scale  for $m$.
$1/f$ noise alone would
appear on the plot as an horizontal line, with amplitude
equal to that of the white noise at 
$m = m_{\rm knee} = f_{\rm knee}/f_{\rm spin}$. As long as $m_{\rm knee}$ is
smaller than a few, the excess $1/f$ noise induces a slight flattening of the
noise curve for low $m$, with negligible total additional noise power.

\section{Other systematics and instrumental effects}

Apart from noise, there are a few effects of instrumental origin which can
be understood much easier on timelines than on maps of the sky. Some of
these effects can be described as those of filters with a known impulse 
response, $h(t)$. The Fourier transform of the impulse response, $H(f)$, 
is known in signal processing as the transfer function of the filter.
For such filters, the filtering theorem states that if a signal $u(t)$ has
a spectrum ${\cal S}_u(t)$, the corresponding filtered signal, 
$u_F(t) = u(t) \star h(t)$, has spectrum ${\cal S}_u(t) |H(f)|^2$. Two 
important instrumental effects can be addressed with this formalism.

\subsection{Time constant of bolometers}

A typical bolometer for CMB anisotropy observations is a temperature-sensitive
resistor heated by incoming radiation and cooled by a heat-conducting 
connection to a sub-kelvin thermal bath. The response of such a system to
an impulse of incident power is:
\be
h(t) = \frac{1}{\tau} \exp(-t/\tau)
\label{eq:h_time_constant}
\ee
The constant $\tau$ is known as the time constant of the bolometer. It is set
by physical characteristics of the bolometer ($\tau = C/G$, where $C$ is the
heat capacity of the bolometer and $G$ the effective conductivity of the link
to the thermal bath). However, other bolometer characteristics, and in 
particular the sensitivity of the bolometer, depend on the same parameters.
Thus, it is crucial, for bolometer optimisation, to understand the effect
of the time constant of the bolometers on the spectrum of the useful
astrophysical (and especially cosmological) signal.

The transfer function $H(f)$ corresponding to the impulse response of
equation \ref{eq:h_time_constant} is:
\be
H(f) = \frac{1}{1+2i \pi f \tau}
\ee
Using the relation between $m$ and frequencies of the signal, $m=f/f_{\rm spin}$, 
the attenuation function $A_m$ on the $\Gamma_m$ corresponding to the effect of the
time constant of the bolometers is:
\be
A_m = |H_m|^2 = \frac{1}{1+(2 \pi f_{\rm spin} \tau)^2 m^2}
\ee
and instead of measuring the quantities $\Gamma_m$, the instrument measures
$A_m\Gamma_m$. 

\subsection{Effect of the sampling frequency}

Another instrumental parameter which needs to be set carefully by designers
of CMB experiments is the sampling frequency. It should be high enough
in order to avoid aliasing, but making it unnecessarily high results
in a very large set of data, with the associated problems of data storage
and telemetry, especially for satellite or long duration ballooning 
experiments.

Again, the effect of the sampling rate can be understood very simply as that
of a filter. If sampling is modelled by a perfect integrator over a period
$T_0$ (which is the inverse of the sampling rate) for obtaining each sample,
the corresponding impulse response is:
\be
h(t) = \frac{1}{T_0} {\rm rect}(T_0),
\ee
where ${\rm rect}(T_0)$ is the function whose value is 1 between $t-T_0/2$
and $t+T_0/2$, and 0 elsewhere. The corresponding transfer function is
\be
H(f) = \frac{\sin(\pi f T_0)}{\pi f T_0}
\ee
and the corresponding attenuation function $A_m$ on the $\Gamma_m$ is:
\be
A_m = |H_m|^2 = \frac{\sin^2(\pi mf_{\rm spin} T_0)}{(\pi mf_{\rm spin}  T_0)^2}
\ee

Attenuation functions for these two effects
are plotted in the bottom panel of Figure \ref{fig:fig_G_m_noise}.

\section{Conclusion}

In this paper, we presented the formalism to project the 
CMB anisotropy predictions and
instumental noise effects on the rings of data. 
Because of the feasibility of  experimental setup for CMB observations
with  circular scans, and simplicity of the data analysis
on such scans, we believe that collecting
CMB anisotropy data in this format is
an interesting option for the next generation of anisotropy experiments.
It provides a natural frame for studying the statistics of CMB anisotropies,
in which many instrumental effects can be modelled and analysed under the most 
natural connection between the time stream of data and the  
spatial distribution of the directions of measurements.
We attempted to illustrate this point  comprehensively  in our 
Figure \ref{fig:fig_G_m_noise}. Given that this comparative plot allows to 
present simultaneously
a natural rendition of a number of factors that affect the attempts to measure
the CMB anisotropy on the celestial rings,
we believe that Fig. \ref{fig:fig_G_m_noise} should provide
significant insights on optimization of experimental setup for future 
CMB experiments.

For the most ambitious CMB programmes such as PLANCK, many such scans will have
to be merged into two-dimensional maps. The circular scan
method described in this paper is a useful intermediate step in the analysis 
of such data in the global analysis of
CMB anisotropies on the celestial sphere.

\section*{Acknowledgments}

Original inspiration for KMG to consider the circular scanning came from
discussions of the UCSB HACME experiment with P. Lubin and M. Seiffert,
whereas JD's interest in the subject was stimulated by discussions with
Fran\c{c}ois-Xavier D\'esert and Jean-Loup Puget over the choice of
an optimal scanning strategy for the DIABOLO 1997 observations. JD would
like to thank Eric Aubourg, Nathalie Palanque-Delabrouille and
Simon Prunet for help and suggestions in numerical computations of $\G_m$'s
for the first draft of this paper.
CMB anisotropy power spectra in the flat CDM models were computed 
using the code CMBFAST (Seljak \& Zaldarriaga, 1996).
JD was supported by a PhD research grant from CNES, the French national space 
agency.
KMG and EH were supported in part by the Danish Research Foundation through its
establishment of the Theoretical Astrophysics Center.





\begin{thebibliography}{99}
\bibitem{a} Bardeen, J.M., \etal, 1986, ApJ, 304, 15
\bibitem{b} Bersanelli \etal, 1996, COBRAS/SAMBA : Report on the Phase A Study
\bibitem{d} Delabrouille, J., 1997, to be published in A\&A supplement series
\bibitem{s} Seljak, U. \& Zaldarriaga, M., 1996, ApJ, 469, 437



\end{thebibliography}
\end{document}